# Non-$d^0$ Electric Dipole in FeO$_5$ bipyramid: a New Resource for Quantum Paraelectrics, Ferroelectrics and Multiferroics

Shi-Peng Shen, Yi-Sheng Chai,* Jun-Zhuang Cong, Pei-Jie Sun, Jun Lu, Li-Qin Yan,

Shou-Guo Wang, and Young Sun†

*Beijing National Laboratory for Condensed Matter Physics, Institute of Physics,*

*Chinese Academy of Sciences, Beijing 100190, China*

*yschai@iphy.ac.cn    †youngsun@iphy.ac.cn

Abstract

Electric polarization in conventional ferroelectric oxides usually involves nonmagnetic transition-metal ions with an empty $d$ shell (the $d^0$ rule). Here we unravel a new mechanism for local electric dipoles based on magnetic Fe$^{3+}$ ($3d^5$) ion violating the $d^0$ rule. The competition between the long-range Coulomb interaction and short-range Pauli repulsion in a FeO$_5$ bipyramid with proper lattice parameters would favor an off-center displacement of Fe$^{3+}$ that induces a local electric dipole. The manipulation of this kind of non-$d^0$ electric dipoles opens up a new route for generating unconventional dielectrics, ferroelectrics, and multiferroics. As a prototype example, we show that the non-$d^0$ electric dipoles in ferrimagnetic hexaferrites (Ba,Sr)Fe$_{12}$O$_{19}$ lead to a new family of magnetic quantum paraelectrics.

The majority of conventional ferroelectrics are transitional metal oxides with a perovskite structure ($ABO_3$) such as $BaTiO_3$ and $(Pb,Zr)TiO_3$. Their electric polarization usually requires an off-center shifting of transition metal ($B$) ions with an empty $d$ shell due to the hybridization between empty $d$ orbital and filled O $2p$ orbitals [1, 2]. This empirical "$d^0$ rule" established by Matthias [3] has severely restrained the family of ferroelectrics and even precludes the coexistence of electric and magnetic orders because magnetism normally requires non-$d^0$ configurations [4]. In order to incorporate ferroelectricity with magnetism, other mechanisms of ferroelectricity without involving the $B$-site off-centering for electric polarization, such as $A$-site lone-pair electrons [5], non-collinear spin configurations [6], charge ordering [7,8], and geometric ferroelectricity [9], have been extensively studied in the past decade.

On the other hand, recent studies suggest that the $d^0$ rule could be broken in certain circumstances. For example, theoretical calculations predicted the possible ferroelectricity in $AMnO_3$ ($A$=Sr, Ca, and Ba) driven by off-centering of the magnetic $Mn^{4+}$ ion in $MnO_6$ octahedron via the second-order Jahn-Teller effect [10-12]. Recently, ferroelectricity was indeed observed in perovskite $Sr_{1-x}Ba_xMnO_3$ ($x \geq 0.45$) single crystals [13]. This progress poses a big challenge to the empirical $d^0$ rule. In this Letter, we demonstrate another intriguing case violating the $d^0$ rule other than the perovskite oxides. It is found that the delicate balance between the long-range Coulomb interaction and short-range Pauli repulsion in a $FeO_5$ bipyramid would favor an off-center displacement of the magnetic $Fe^{3+}$ ions, which directly induces a local electric dipole.

This type of non-$d^0$ electric dipole provides a new resource for unconventional dielectrics, ferroelectrics, and multiferroics.

The principle unraveled in this work initially originates from the understanding of the abnormal dielectric behaviors in the M-type hexaferrites (Ba,Sr)Fe$_{12}$O$_{19}$. Hexaferrites are iron oxides with hexagonal structures. Depending on their chemical formula and crystal structures, hexaferrites can be classified into several types [14,15]: M-type (Ba,Sr)Fe$_{12}$O$_{19}$, Y-type (Ba,Sr)$_2$*Me*$_2$Fe$_{12}$O$_{22}$, W-type (Ba,Sr)*Me*$_2$Fe$_{16}$O$_{27}$, X-type (Ba,Sr)$_2$*Me*$_2$Fe$_{28}$O$_{46}$, Z-type (Ba,Sr)$_3$*Me*$_2$Fe$_{24}$O$_{41}$, and U-type (Ba,Sr)$_4$*Me*$_2$Fe$_{36}$O$_{60}$, where *Me* = a bivalent metal ion. As shown in Fig. 1, the structures of hexaferrites can be described by a periodically stacking sequence of three basic building blocks (S, R, and T blocks) along $c$ axis. The Fe$^{3+}$ ions occupy three different kinds of sites: octahedral, tetrahedral, and bipyramidal sites. In particular, the FeO$_5$ bipyramids only exist in the middle of R blocks where the equatorial plane of the bipyramid is a mirror plane. Among all the hexaferrites, the M-type hexaferrites have the simplest crystal structure, consisting of alternate stacks of S and R blocks along $c$ axis. Both BaFe$_{12}$O$_{19}$ and SrFe$_{12}$O$_{19}$ have a collinear ferrimagnetic ordering along $c$ axis, with a Neel temperature of ~ 720 [16] and 737 K [17], respectively. This spin configuration persists down to the lowest temperature without any magnetic transition.

We have prepared various single-crystal and polycrystalline samples of hexaferrites to investigate their dielectric properties. The details of sample preparation and characterization are given in the Supplementary Material [18]. The dielectric permittivity was measured using an Agilent 4980A LCR meter at 1 MHz in a Cryogen-free Superconducting Magnet System (Oxford Instruments, TeslatronPT). Specific heat measurement was performed in a Quantum Design Physical Properties Measurement System.

As shown in Fig. 2(a), the $c$-axis dielectric permittivity $\varepsilon_c$ ($T$) of $BaFe_{12}O_{19}$ increases steadily with decreasing temperature and remains nearly constant below ~ 5.5 K. This saturation behavior in dielectric permittivity resembles well that of quantum paraelectrics such as $SrTiO_3$ [19], $CaTiO_3$ [20], and $KTaO_3$ [21], where the onset of ferroelectric order is suppressed by quantum fluctuations. While the Curie-Weiss law describes well the paraelectric $\varepsilon$ ($T$) at high temperatures, it fails in the region of a quantum paraelectric state. Instead, the mean-field Barret formula [22] is often used to describe the $\varepsilon$ ($T$) of quantum paraelectrics in the entire temperature region:

$$\varepsilon = \varepsilon_0 + \frac{M}{\left(\frac{1}{2}T_1\right)\coth\left(\frac{T_1}{2T}\right) - T_0} \tag{1}$$

where $\varepsilon_0$ is a constant, $T_0$ is proportional to the effective dipole-dipole coupling constant and the positive and negative values correspond to ferro and antiferroelectric interactions, respectively. $T_1$ represents the tunneling integral and is a dividing temperature between the low temperature region where quantum fluctuation is important and the high temperature region where quantum effect is negligible. $M=n\mu^2/k_B$, where $n$ is the density of dipoles and $\mu$ denotes the local dipolar moment.

We find that the $\varepsilon_c$ ($T$) of $BaFe_{12}O_{19}$ can be well fitted by Eq. (1) in the entire temperature region. The parameters obtained from the fitting curve are listed in Table I. The negative value of $T_0$ = -11.7 K implies an antiferroelectric coupling between electric dipoles. The value of $T_1$ = 54.9 K is comparable to that of SrTiO$_3$, but $M$ = 452 K is two orders of magnitude smaller than that of SrTiO$_3$. We note that this quantum paraelectric behavior is observable only along $c$ axis. As shown in Fig. 2(b) for comparison, the *ab*-plane dielectric permittivity measured along the [100] direction decreases monotonically with decreasing temperature. This result suggests that the electric dipoles in BaFe$_{12}$O$_{19}$ are along $c$ axis. To further verify that the saturation of $\varepsilon_c$ at low temperatures has nothing to do with any phase transition, we measured the specific heat $C_P$ of BaFe$_{12}$O$_{19}$ down to 2 K. No anomaly due to a phase transition could be detected (see the Supplementary Material [18]). Therefore, BaFe$_{12}$O$_{19}$ is a distinctive quantum paraelectric with a hexagonal structure, in contrast to those well-known perovskite quantum paraelectrics.

We then checked the consequence of Sr doping on the quantum paraelectricity. As shown in Figs. 2(c) and 2(d), both Ba$_{0.5}$Sr$_{0.5}$Fe$_{12}$O$_{19}$ and SrFe$_{12}$O$_{19}$ also exhibit quantum paraelectric behavior in the $\varepsilon_c$ ($T$). Especially, the plateau in $\varepsilon_c$ ($T$) extends to higher temperatures with higher Sr content, indicating that Sr doping actually enhances quantum fluctuations. The fitting parameters of Eq. (1) to both $\varepsilon_c$ ($T$) curves indeed give a larger $T_1$ in Ba$_{0.5}$Sr$_{0.5}$Fe$_{12}$O$_{19}$ than that in BaFe$_{12}$O$_{19}$ (Table I). Besides, Sr doping reduces the magnitude of both $T_0$ and $M$, indicating a weaker effective dipole-dipole coupling and a smaller local dipole moment in Ba$_{0.5}$Sr$_{0.5}$Fe$_{12}$O$_{19}$. Note that, there is a relatively large $\varepsilon_c$ ($T$) background above 100 K in $SrFe_{12}O_{19}$ due to the rise of conductivity at high temperatures. Therefore, the fitting parameters of $T_0$ and $T_1$ for $SrFe_{12}O_{19}$ are not reliable while the parameter $M$ is not so sensitive to the high temperature background to lose physical meaning. In this sense, we are safe to say that the local dipole moment is suppressed by Sr doping.

The above results reveal that all the members of $(Ba,Sr)Fe_{12}O_{19}$ are quantum paraelectrics. As they also have a ferrimagnetic ground state, $(Ba,Sr)Fe_{12}O_{19}$ represents a new family of magnetic quantum paraelectrics. Since the first discovery of quantum paraelectricity in $SrTiO_3$ by Müller *et al.* in 1979 [19], magnetic quantum paraelectrics where quantum paraelectricity coexists with magnetic ordering have been rarely observed. The only known example so far is $EuTiO_3$ that exhibits quantum paraelectricity along with an antiferromagnetic order [23]. The $(Ba,Sr)Fe_{12}O_{19}$ family is the first example of ferrimagnetic quantum paraelectrics ever known.

To understand the physical mechanism underlying the magnetic quantum paraelectricity in $(Ba,Sr)Fe_{12}O_{19}$, we need to find out the origin of local electric dipoles first. As seen in Fig. 1, the structure of $(Ba,Sr)Fe_{12}O_{19}$ consists of alternating R and S blocks. The $FeO_5$ bipyramids residing in the middle of the R blocks, if in the centrosymmetric structure with space group $P6_3/mmc$, would not generate local electric dipoles. However, some experiments [24,25] have suggested the existence of off-equatorial displacements of $Fe^{3+}$ at Wyckoff position of 2b site, which results in two adjacent Wyckoff positions of 4e sites with a lowered symmetry. If the off-equatorial displaced $Fe^{3+}$ at the 4e sites has a lower energy than the high symmetric 2b sites, a local electric dipole *P* along *c* axis would be favored, as illustrated in Fig. 3(a). Therefore, the highly anisotropic displacements of $Fe^{3+}$ at 4e sites could be the origin of the local dipoles we are looking for. Moreover, the parameter *M* is found to decrease systematically with Sr doping (Table I). From the fitted *M* values and the nominal density of 4e sites, $n = 2.86 \times 10^{27}$ m$^{-3}$, the local dipole moments $\mu$ = 0.68, 0.50, and 0.09 *e*Å are obtained for $BaFe_{12}O_{19}$, $Ba_{0.5}Sr_{0.5}Fe_{12}O_{19}$, and $SrFe_{12}O_{19}$, respectively. The decreasing dipole moment with increasing Sr content suggests a smaller off-equatorial displacements for the 4e sites, which is consistent with the x-ray refinement results [24,26] that the 4e-4e distance at room temperature decreases from 0.34(1) Å to 0.194(10) Å from $BaFe_{12}O_{19}$ to $SrFe_{12}O_{19}$. Mössbauer study [27] also revealed a similar tendency at 4.2 K that the 4e-4e distances are 0.176(5) and 0.133 for $BaFe_{12}O_{19}$ and $SrFe_{12}O_{19}$, respectively.

To give a more convincing assessment to the origin of local dipoles, we did theoretical simulations on the local potential energy profiles along *c* axis within the $FeO_5$ bipyramid for $(Ba,Sr)Fe_{12}O_{19}$. We adopted a phenomenological local potential energy of the following form for the bipyramid [28]:

$$U_{tot}(z) = U_{Coulomb}(z) + U_{repulsion}(z) = -\frac{3 \times 6e^2}{\sqrt{r_0^2 + z^2}} - \frac{6e^2}{r_1 + z} - \frac{6e^2}{r_1 - z} +$$

$$3\beta c_{+-}e^{(r_+ + r_- - \sqrt{r_0^2 + z^2})/\rho} + \beta c_{+-}e^{(r_+ + r_- - (r_1+z))/\rho} + \beta c_{+-}e^{(r_+ + r_- - (r_1-z))/\rho} \quad (2)$$

where *z* is the *c*-axis $Fe^{3+}$ displacement away from the 2b site. The sum of first three terms corresponds to the Coulomb potential $U_{Coulomb}$ between $Fe^{3+}$ and $O^{2-}$ and the sum of the rest three terms represents the short-range Pauli repulsion potential $U_{repulsion}$. (Ba,Sr)Fe$_{12}$O$_{19}$ is assumed to be a pure ionic crystal that O ion has -2e charge and Fe ion has +3e charge, $e$ is the electron charge. $r_0$ and $r_1$ are the in-plane and out of plane Fe-O distances in the bipyramid for 2b site, obtained from ref. [29] and Table I. $\beta$ is a constant (taken to be $1.35 \times 10^{-19}$ J), $c_{+-} = 1$ is Pauling's valence factor, and $\rho = 0.315$ Å is a parameter derived from the FeO (ref. [28]). $r_+ = 1.4$ Å and $r_- = 0.58$ Å are the ionic radii of Fe$^{3+}$ with coordinate number of 5 and O$^{2-}$ with coordinate number of 6, respectively, from ref. [30].

The calculated results are shown in Figs. 3(b)-3(d). Only the $U_{repulsion}$ terms show the double-well potential feature near $z = 0$ for the parameters we used. The double well in $U_{repulsion}$ terms becomes smaller and closer with Sr doping while the $U_{Coulomb}$ terms are largely invariant near $z = 0$. Their summation $U_{tot}$ remains the double-well feature, indicating the existence of local electric dipoles in (Ba,Sr)Fe$_{12}$O$_{19}$. In particular, the 4e-4e distances estimated from the simulation decreases systematically from 0.271 to 0.142 Å by substituting Ba with Sr, which is well consistent with the decreasing tendency of local dipole by Sr doping found in our experiments. Based on above calculations, the short-range Pauli repulsions instead of the Coulomb forces favor the off-equatorial arrangement of Fe$^{3+}$ ion. This is in marked contrast to the case in $ABO_3$ perovskite ferroelectrics where the long-range Coulomb forces favor the ferroelectric state and short-range repulsions do not. It is very likely due to the unique FeO$_5$ bipyramid structure as well as the half-filled configuration of Fe$^{3+}$ ($3d^5$).

To provide further insight for the lattice requirements to have the local electric dipoles in FeO$_5$ bipyramid, we analyzed the extreme conditions of each potential function at $z = 0$. For Coulomb term, to favor off-center displacement of Fe$^{3+}$ ion, it should have $\partial^2 U_{Coulomb}/\partial z^2|_{z=0} < 0$. One can obtain the corresponding structure requirement: $r_1/r_0 < (4/3)^{1/3} \approx 1.1$. This condition is always unsatisfied for (Ba,Sr)Fe$_{12}$O$_{19}$ since $r_1/r_0 > 1.23$ for all the members. For the repulsion term, to have the double-well feature, it should have $\partial^2 U_{repulsion}/\partial z^2|_{z=0} < 0$. One then obtains the lattice criteria: $e^{r_1/\rho} > 2/3(r_0/\rho\, e^{r_0/\rho})$. The satisfaction of this criterion depends on the relative magnitude of $r_1$ and $r_0$ to $\rho$, and it happens to be the case for (Ba,Sr)Fe$_{12}$O$_{19}$. By combining the conditions of both potential terms, we obtained the phase diagram in the $r_1/\rho$ vs. $r_0/\rho$ plot, as shown in Fig. 3(e). (Ba,Sr)Fe$_{12}$O$_{19}$ are found to lie just outside the nonpolar region.

The produced double-well potentials can only guarantee the existence of local dipoles. Nevertheless, to observe quantum paraelectricity, quantum fluctuations must overwhelm the coupling between local dipoles to prevent a ferro or antiferroelectric ordering. A recent first-principle calculation on BaFe$_{12}$O$_{19}$ predicted a frustrated antiferroelectric ground state below 3 K [31]. On the contrary, we neither observe in experiments any ferro/antiferroelectric phase transition down to 1.5 K, nor found any trace of long-range ordering in the specific heat data above 2 K. These experimental facts indicate that the previous first-principle calculation could have either ignored the quantum fluctuations or overestimated the dipole-dipole coupling in BaFe$_{12}$O$_{19}$. Since the local dipoles are from the FeO$_5$ bipyramid, the density of dipoles is proportional to the number of FeO$_5$ bipyramids in the whole lattice. Due to the layered structures, the local dipoles in FeO$_5$ bipyramids only exist in the R blocks and are separated by the S blocks along *c* axis. In this case, the coupling between these diluted small dipoles would be quite weak, especially along *c* axis. Meanwhile, the double-well potentials in (Ba,Sr)Fe$_{12}$O$_{19}$ are very shallow and therefore can be easily disturbed by quantum fluctuations. From above discussions, we conclude that the quantum fluctuations between two opposite off-equatorial configurations overwhelm the weak dipole-dipole coupling and prevent the long-range dipole ordering down to the lowest temperature.

If the local dipoles are really related to the FeO$_5$ bipyramids within the R blocks, one may expect a similar dielectric behavior in other hexaferrites containing the R blocks. To clarify this point, we further studied a series of available hexaferrites, including Y-type (BaSrCoZnFe$_{11}$AlO$_{22}$) and Z-type (Ba$_{1.5}$Sr$_{1.5}$Co$_2$Fe$_{24}$O$_{41}$) single crystals, and W-type (BaCo$_2$Fe$_{16}$O$_{27}$) ceramics. For Y and Z-type hexaferrites, a constant in-plane magnetic field of 5 kOe is applied to stabilize the transverse cone spin configuration below 150 K during the measurements [32,33]. As shown in Figs. 4(a) and 4(b), the W- and Z-type hexaferrites that contain the R blocks indeed shows a sign of quantum paraelectricity, with a similar saturating behavior to that in (Ba,Sr)Fe$_{12}$O$_{19}$. In strong contrast, the Y-type hexaferrite does not show such a saturation behavior. This is exactly consistent with our expectation because Y-type hexaferrites do not contain the FeO$_5$ bipyramids.

In conclusion, we have disclosed a new mechanism for local electric dipoles based on the FeO$_5$ bipyramid. As a prototype consequence, this non-$d^0$ electric dipoles in hexagonal ferrites (Ba,Sr)Fe$_{12}$O$_{19}$ leads to a completely new family of magnetic quantum paraelectrics. More importantly, the significance of the FeO$_5$ bipyramid goes well beyond quantum paraelectrics only. In principle, a variety of exotic dielectric materials could be expected on the basis of the non-$d^0$ electric dipole. The amplitude of the electric dipoles and the coupling between them may be enhanced by manipulating the lattice parameters and the density of the bipyramid unit in a designed crystal structure, which may eventually yield unconventional non-$d^0$ ferroelectrics. Furthermore, the coexistence of both electric dipole and magnetic moment in the FeO$_5$ bipyramid provides a promising playground for magnetoelectric multiferroics. These topics should deserve further studies in the future.

This work was supported by the Natural Science Foundation of China under Grant Nos. 11227405 and 11374347, the National Key Basic Research Program of China under Grant No. 2011CB921801, and the China-Israel joint project under Grant No. 2013DFG13020.

**References**

[1] R. E. Cohen, Nature **358**, 136 (1992).

[2] R. E. Cohen and H. Krakauer, Ferroelectrics **136**, 65 (1992).

[3] B. T. Matthias, Phys. Rev. **75**, 1771 (1949).

[4] N. A. Hill, J. Phys. Chem. B **104**, 6694 (2000).

[5] R. Seshadri and N. A. Hill, Chem. Mater. **13**, 2892 (2001).

[6] S.-W. Cheong and M. Mostovoy, Nature Mater. **6**, 13 (2007).

[7] N. Ikeda *et al*., Nature **436**, 1136 (2005).

[8] Y. Sun, L.-Q. Yan, and J.-Z. Cong, Sci. China-Phys. Mech. Astron. **56**, 222 (2013).

[9] B. B. van Aken *et al*., Nature Mater. **3,** 164 (2004).

[10] S. Bhattacharjee *et al*., Phys. Rev. Lett. **102,** 117602 (2009).

[11] J. M. Rondinelli *et al.*, Phys. Rev. B **79**, 205119 (2009).

[12] J. H. Lee and K. M. Rabe, Phys. Rev. Lett. **104**, 207204 (2010).

[13] H. Sakai *et al.*, Phys. Rev. Lett. **107**, 137601 (2011).

[14] P. B. Braun, *Phillips Res. Rep.* **12**, 491 (1957).

[15] J. A. Kohn, D. W. Eckart, and C. F. Jr Cook, *Science* **172**, 519 (1971).

[16] J. Smit and H. P. J. Wijn, *Ferrites* (Phillips Technical Library, 1959).

[17] Z. F. Zi *et al.*, J. Magn. Magn. Mater. **320**, 2746 (2008).

[18] See Supplemental Material for the details of sample preparation, characterization, and heat capacity.

[19] K. A. Müller and H. Burkhard, Phys. Rev. B **19**, 3593 (1979).

[20] V. V. Lemanov, A. V. Sotnikov, and E. P. Smirnova, *Solid State Commun.* **110**, 611 (1999).

[21] A. R. Akbarzadeh *et al.*, Phys. Rev. B **70**, 054103 (2004).

[22] J. H. Barret, Phys. Rev. **86**, 118120 (1952).

[23] T. Katsufuji and H. Takagi, Phys. Rev. B **64**, 054415 (2001).

[24] X. Obradors *et al.*, *J. Solid State Chem.* **56**, 171 (1985).

[25] J. G. Rensen, and J. G. van Wieringen, *Solid State Commun.* **7**, 1139 (1969).

[26] X. Obradors *et al.*, *Solid State Chem.* **72**, 218 (1988).

[27] G. Albanses, A. Deriu, and D. Cabrini, *Hyperfine Interactions* **70**, 1087 (1992).

[28] S. P. Marshall and J. B. Sokoloff, Phys. Rev. B **44**, 619 (1991).

[29] O. P. Aleshko-Ozhevskii, M. K. Faek, and I. I. Yamzin, *Soviet Physics-Cryst.* **14**, 447 (1969).

[30] R. D. Shannon, *Acta Cryst.* A**32**, 751 (1976).

[31] P. S. Wang and H. J. Xiang, Phys. Rev. X **4**, 011035 (2014).

[32] S.-P. Shen *et al.*, Appl. Phys. Lett. **104**, 032905 (2014).

[33] M. Soda *et al.*, Phys. Rev. Lett. **106**, 087201 (2011).

**Figure captions**

FIG. 1 (color online). Crystal structures of M, Y, W, and Z-type hexaferrites.

FIG. 2 (color online). The *c*-axis dielectric permittivity $\varepsilon_c$ as a function of temperature for (a) $BaFe_{12}O_{19}$, (c) $Ba_{0.5}Sr_{0.5}Fe_{12}O_{19}$, and (d) $SrFe_{12}O_{19}$. (b) The *ab*-plane dielectric permittivity $\varepsilon_{ab}$ for $BaFe_{12}O_{19}$. The black solid lines are the fitting curves to Eq. (1). The insets are the same plots with a logarithmic scale in temperature.

FIG. 3 (color online). (a) Schematic illustration of $Fe^{3+}$ off-equatorial displacements in the $FeO_5$ bipyramid. The up and down displacements correspond to two 4e sites with opposite dipoles and local minimums in energy potential. The calculated energy potentials from Eq. (2) for (b) $U_{Coulomb}$, (c) $U_{repulsion}$, and (d) the sum $U_{tot}$ as a function of off-equatorial $Fe^{3+}$ displacement $z$ in $(Ba,Sr)Fe_{12}O_{19}$. (e) Phase diagram to allow the double-well potential. In the central yellow region, the double-well potential is not allowed. In the left corner, it must appear. In the rest regions, it is possible due to the competition between the Coulomb attraction and Pauli repulsion.

FIG. 4 (color online). The temperature dependent dielectric permittivity of (a) the W-type hexaferrite $BaCo_2Fe_{16}O_{27}$ ceramic, (b) the Z-type hexaferrite $Ba_{1.5}Sr_{1.5}Co_2Fe_{24}O_{41}$ single crystal along *c* axis, and (c) the Y-type hexaferrite $BaSrCoZnFe_{11}AlO_{22}$ single crystal along *c* axis.

**Table I.** The fitting parameters $T_0$, $T_1$ and $M$ to Eq. (1). The values in bold are not reliable. The Fe-O bond lengths in the FeO$_5$ bipyramid for BaFe$_{12}$O$_{19}$ and SrFe$_{12}$O$_{19}$ at room temperature are derived from ref. [29]. The bond lengths of Ba$_{0.5}$Sr$_{0.5}$Fe$_{12}$O$_{19}$ are linearly extrapolated from that of the end members.

| Ba$_{1-x}$Sr$_x$Fe$_{12}$O$_{19}$ | x = 0.0 | x = 0.5 | x = 1.0 |
|---|---|---|---|
| $T_0$ (K) | -11.7 | -9.7 | **53.1** |
| $T_1$ (K) | 54.9 | 67.6 | **130.6** |
| $M$ (K) | 452 | 224 | 7.4 |
| Fe-O$_{ap}$ (Å) | 2.325 | 2.315 | 2.305 |
| Fe-O$_{ab}$ (Å) | 1.870 | 1.8665 | 1.863 |

Figure 1

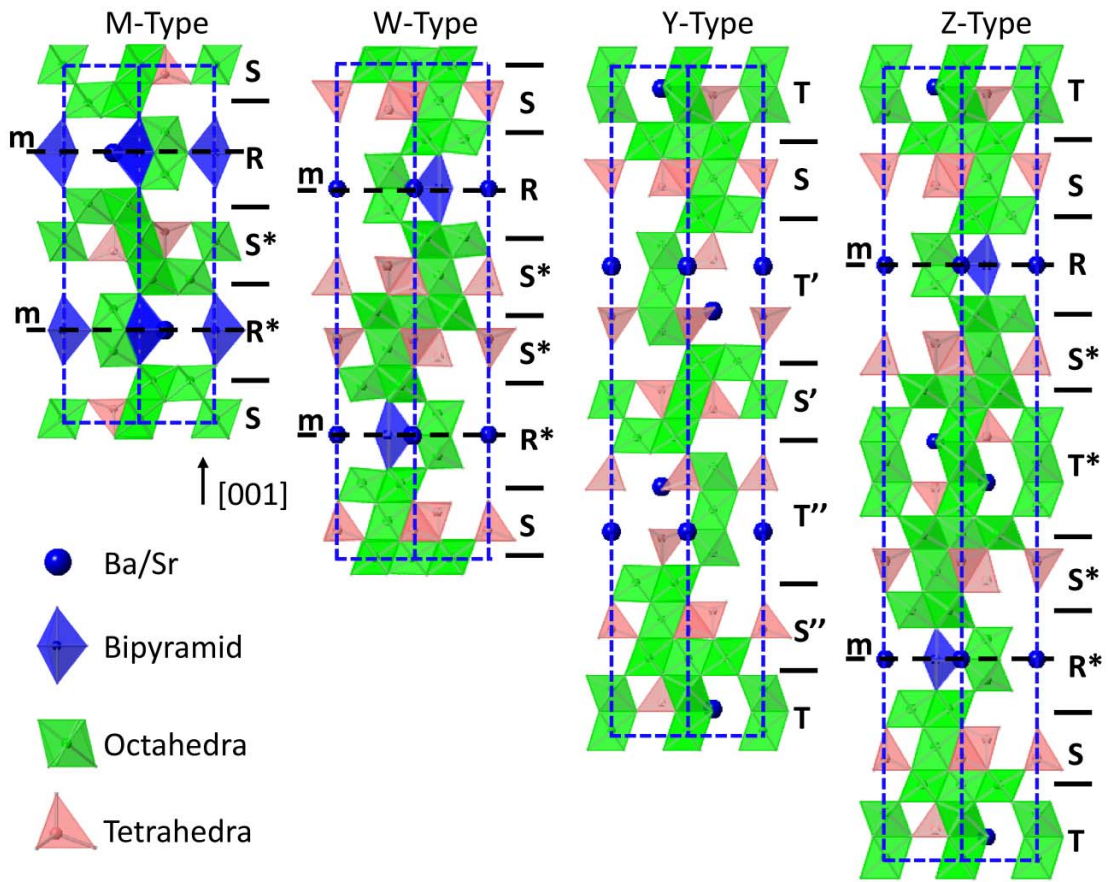

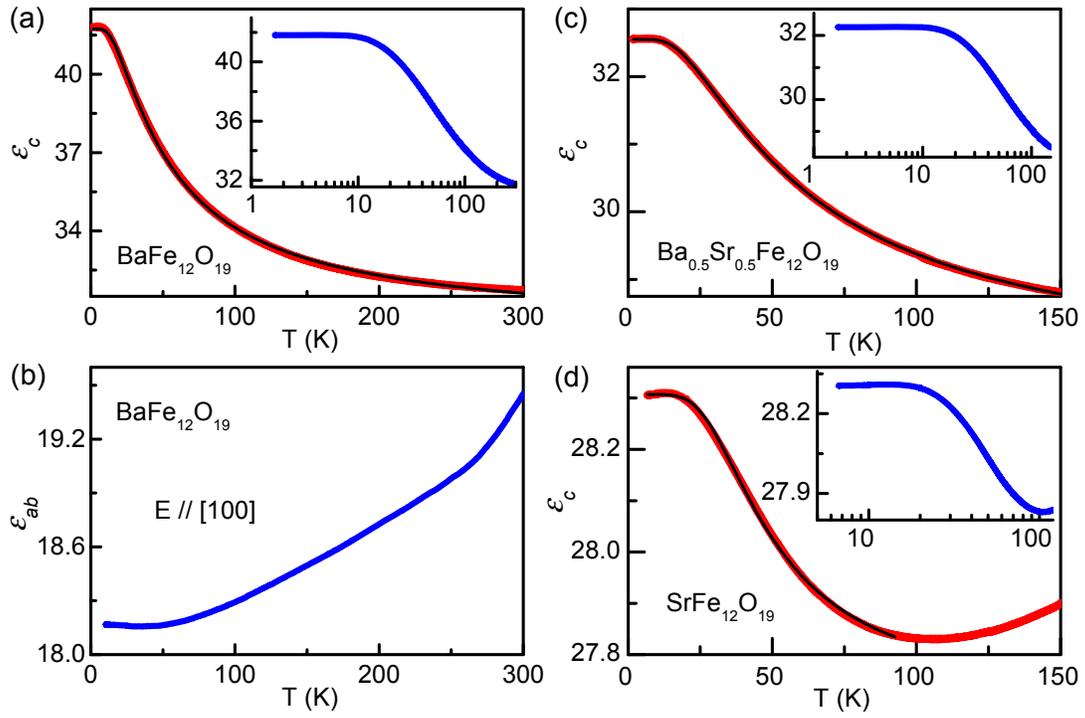

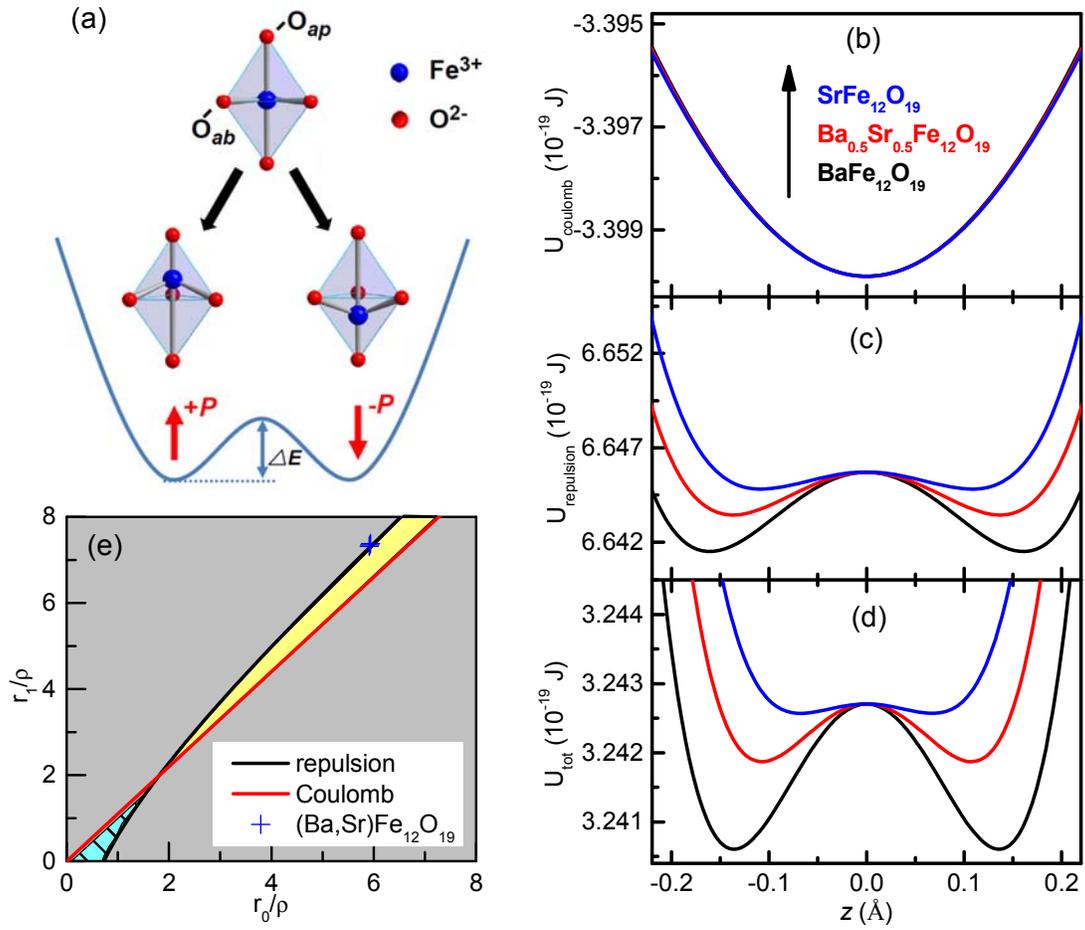

Figure 4

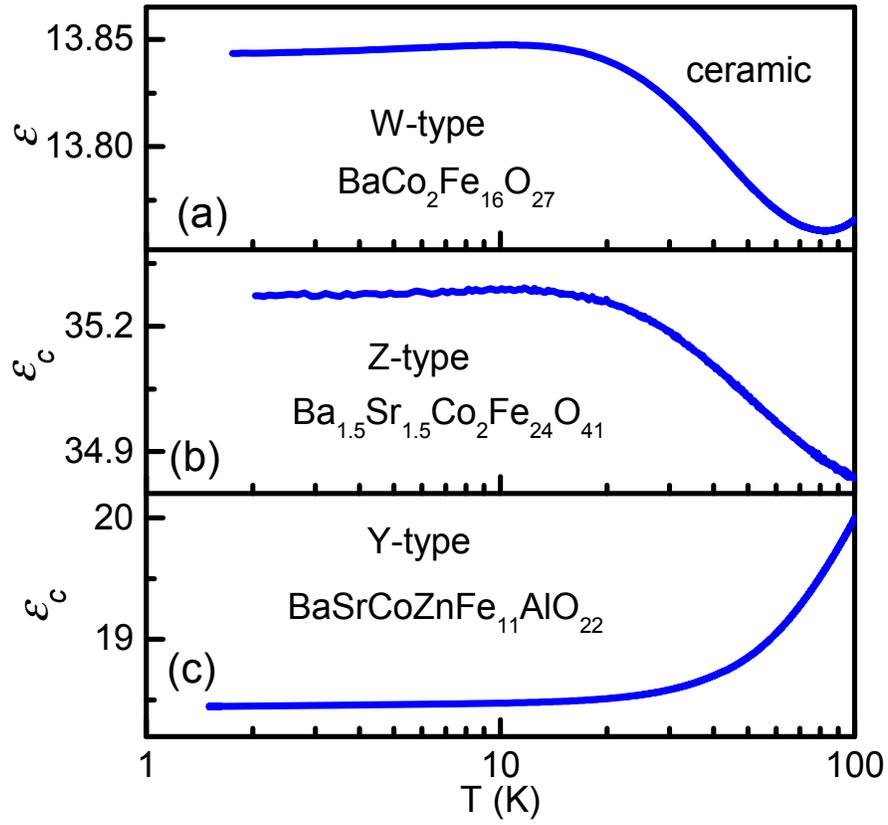